\documentclass[conference]{IEEEtran}

\usepackage{amsmath,amsopn,amsthm,wasysym,mathtools}
\usepackage{float}
\usepackage{subcaption}
\usepackage{endnotes,microtype,xspace,graphicx,fancyvrb,multirow}
\usepackage{booktabs}
\usepackage{enumitem}
\usepackage[labelfont=bf,font=small,skip=5pt]{caption}
\usepackage[ruled,linesnumbered]{algorithm2e}
\SetKwInput{kwInit}{Init}
\SetKwComment{Comment}{\textnormal{/*} }{ \textnormal{*/}}

\usepackage{fp}
\usepackage{siunitx}

\usepackage[acronym,nohypertypes={acronym,notation}]{glossaries}

\usepackage{balance}

\usepackage{tikz}
\usetikzlibrary{calc}
\usetikzlibrary{positioning}
\usetikzlibrary{fit}
\usetikzlibrary{shapes}
\usetikzlibrary{trees}
\usetikzlibrary{arrows}
\usetikzlibrary{overlay-beamer-styles}
\usepackage{xcolor}

\usepackage{changepage}

\usepackage{balance}
\usepackage{multicol}
\usepackage{etoolbox}

\usepackage{comment}
\usepackage[]{hyperref}
\hypersetup{                    
  colorlinks,
  linkcolor={green!80!black},
  citecolor={red!70!black},
  urlcolor={blue!70!black}
}

\usepackage[square,comma,numbers,sort&compress]{natbib}
\usepackage{tablefootnote}
\usepackage{textcomp}
\usepackage{mdframed}

\usepackage{xurl}

\newacronym{DoS}{DoS}{denial-of-service}
\newcommand{\type}{\mbox{\textsc{SeType}}\xspace}
\newcommand{\sys}{\mbox{\textsc{PySeType}}\xspace}
\newcommand{\numbug}{XYZ\xspace}

\newcommand{\nvul}{12\xspace}
\newcommand{\nprojects}{103\xspace}
\newcommand{\detectrate}{87\%\xspace}
\newcommand{\acrate}{88\%\xspace}

\newcommand{\MX}[1]{\textcolor{magenta}{MX: #1}}

\newcommand{\cc}[1]{\mbox{\smaller[0.5]\texttt{#1}}}

\fvset{fontsize=\scriptsize,xleftmargin=8pt,numbers=left,numbersep=5pt}

\input{code/fmt}

\def\Snospace~{Section {}}

\usepackage{amssymb}
\usepackage{pifont}

\newif\ifdraft\drafttrue
\newif\ifnotes\notestrue
\ifdraft\else\notesfalse\fi

\input{glyphtounicode}
\newcolumntype{R}[1]{>{\raggedleft\let\newline\\\arraybackslash\hspace{0pt}}p{#1}}

\newcommand{\squishlist}{
\begin{itemize}[noitemsep,nolistsep]
  \setlength{\itemsep}{-0pt}
}
\newcommand{\squishend}{
  \end{itemize}
}

\usepackage{tikz}
\newcommand*\WC[1]{%
\begin{tikzpicture}[baseline=(C.base)]
\node[draw,circle,inner sep=0.2pt](C) {#1};
\end{tikzpicture}}

\usepackage{xstring}
\newcommand{\PP}[1]{
\vspace{2px}
\noindent{\bf \IfEndWith{#1}{.}{#1}{#1.}}
}

\definecolor{black}{rgb}{0,0,0}
\definecolor{Orange}{rgb}{1,0.5,0}
\definecolor{Red}{rgb}{1,0,0}
\definecolor{Green}{rgb}{0,0.8,0.5}
\definecolor{Purple}{rgb}{0.75,0,1}
\definecolor{babypink}{rgb}{0.96, 0.76, 0.76}
\definecolor{cerulean}{rgb}{0.0, 0.48, 0.65}
\definecolor{azure}{rgb}{0,0.49,1}
\definecolor{periwinkle}{rgb}{0.8, 0.8, 1.0}
\definecolor{Pink}{RGB}{255, 102, 204}
\definecolor{electriccyan}{rgb}{0.0, 1.0, 1.0}
\definecolor{dodgerblue}{rgb}{0.12, 0.56, 1.0}

\newcommand{\boxbeg}{
\vspace{2px}
\noindent\begin{tabular}{|l|}\hline
\begin{minipage}{3.2in}
\vspace{2px}
\noindent
}

\newcommand{\boxend}{
\vspace{2px}
\end{minipage}\\ \hline
\end{tabular}
\vspace{-10pt}
}

  {\list{}{\leftmargin=#1\rightmargin=#1}\item[]}%
  {\endlist}

\newcounter{finding}

\newif\ifarxiv
\arxivtrue   

\newcommand{\twovers}[2]{\ifarxiv #2\else #1\fi}

\begin{document}

\date{}
\title{Finding Vulnerabilities via LLM-Augmented Semantics-Aware Type-Checking}

\twovers{
\author{
    \IEEEauthorblockN{Anonymous Author(s)}
    \IEEEauthorblockA{Anonymouse Institution\\
    annoymouse@annoymous.com}
}}{
\author{
    \IEEEauthorblockN{Ruizhe Wang}\\
    \IEEEauthorblockA{University of Waterloo}
\and
    \IEEEauthorblockN{Meng Xu}\\
    \IEEEauthorblockA{University of Waterloo}
\and
    \IEEEauthorblockN{N. Asokan}\\
    \IEEEauthorblockA{University of Waterloo}
}
}

\maketitle
\begin{abstract}
Vulnerability detection via static analysis traditionally relies on security experts encoding insecure coding patterns into algorithmic rules. 
However, this approach often focuses on syntactic patterns and overlooks deeper semantic information in the code, such as the meanings of variable and function names. 
As software systems grow more complex, modeling vulnerabilities using only syntactic rules becomes increasingly challenging.

In this paper, we propose a semantics-aware approach to detecting software vulnerabilities. 
We present \type, a semantics-aware type system that can be derived directly from source code based solely on the meanings of symbols and expressions in natural language. 
In the \type type system, both type inference and checking are performed by Large Language Models (LLMs), and a failed type check indicates a potential vulnerability.

We prototype \sys to demonstrate the feasibility of \type for detecting vulnerabilities in Python web applications. 
Our evaluation on real-world applications achieves \detectrate detection precision and \acrate detection accuracy. 
Using \sys, we identified 15 potential zero-day vulnerabilities, nine of which were confirmed by developers.
\end{abstract}

\section{Introduction}

Static analysis is a widely adopted technique for identifying potential software vulnerabilities.
Instead of executing the program,
static analysis tools examine the code through information such as syntax, structure (e.g., control flow), and data flow,
with the goal of finding insecure code patterns
through a set of predefined rules~\cite{zhou2024comparison}.

To systematically identify vulnerabilities, static analysis tools must reason about the security implications of operations and data flows in a program.
For example, dereferencing a \cc{@Nullable} pointer might lead to a \cc{null-pointer} bug.
This resembles type systems in programming languages: static analysis can be conceptualized as a form of type checking that aims to
\emph{assign} type labels to variables and expressions and
\emph{check} whether certain type labels may participate in a given operation.
In this context, the checking policy enforced by the static analyzer would classify such a dereference as a type-mismatch vulnerability.

However,
while \emph{describing} a vulnerability in natural language is often straightforward,
precisely \emph{capturing} its semantics through a set of \textit{syntactic} rules
built upon abstract syntax trees (ASTs),
control-flow graphs (CFGs), or more generally
program-dependence graphs (PDGs)~\cite{codeql,bandit,semgrep}
is challenging.
Indeed, the problem of accurately capturing program semantics is known to be undecidable~\cite{rice_classes_1953}, meaning that no algorithm can determine the correctness of all possible programs.
As a result, existing static-analysis tools often rely on manually crafted rules to approximate semantics
and struggle to balance detection rates with false alarms.
For example, CodeQL builds nearly 500 handwritten rules~\cite{avgustinov_ql_2016,codeql}
yet identifies less than 25\% of bugs in practice~\cite{li_iris_2025}.
On the other hand,
improving bug detection rate often requires
adopting more aggressive rules, which inevitably leads to high false positive rates~\cite{guo2023mitigating,lee2019classifying}
that overwhelm developers, and erode their trust in the tool.
These limitations largely stem from the difficulty of translating program \textit{semantics}
into manually crafted \textit{syntactic} rules.
Ideally, a static-analysis tool would operate directly on program \textit{semantics},
precisely identifying bugs while minimizing false alarms.

To address this gap, we propose a semantics-aware approach to statically detecting software vulnerabilities.
We introduce \type, a type system that operates on program semantics rather than syntactic rules.
In \type, every variable and expression is assigned a \textit{semantic type} that captures its intended meaning and behavior.
Borrowing from the bidirectional type system~\cite{dunfield2021bidirectional}, we design each operation to both \textit{check} and \textit{infer} the \textit{semantic types} of its operands and results.
If the entire program type-checks, we consider it free of the vulnerabilities detectable by \type.

To illustrate,
we present a path traversal vulnerability (CWE-22) in~\autoref{code:semantics-bug-exp}.
\type checking starts from the \cc{review\_paper} endpoint.
Following the program execution flow,
it first infers \cc{pdf\_path} as ``\cc{user input}'' type (line 5) and then 
\cc{abs\_pdf\_path} as ``\cc{file path from user}'' type (line 11).
This ``\cc{file path from user}'' type carries over to the
\cc{pdf\_path} argument (line 20)
through the function call (line 16) and
participate in the \cc{to\_markdown} operation from the \cc{pymupdf4llm} package (line 21),
which is an agnostic file parser that reads the file at the given path, regardless of its source.
As such, the \cc{review} function expects a properly sanitized path argument, and 
an argument of the ``\cc{file path from user}'' type
violates the typing rules and causes \type to raise a type error.

\begin{figure}[t]
    \centering
    \small
    \hfill\begin{minipage}[c]{\linewidth}\vspace{0.5em}\input{code/example_rdagent}\end{minipage}\vspace{0.5em}
    \caption{Example path traversal vulnerability adopted from (CVE-2025-55149)~\cite{rdagent}. The vulnerability arises from the unsanitized use of user input \cc{pdf\_path} to access the file system in line 21.}
    \label{code:semantics-bug-exp}
\end{figure}

In the above analysis,
the types inferred and checked by \type (e.g., ``\cc{file path from user}'') are not conventional programming types (e.g., \cc{int} and \cc{string}), 
whose meanings are defined by the language.
Instead, they are natural-language descriptions of the semantics of variables
and expressions, generated from the source code text rather than from ASTs, CFGs, or PDGs.
Similarly, the \type checking described above is not a mechanical algorithm. It relies on the natural-language description of the \emph{path traversal} vulnerability and is a fuzzy (non-deterministic) check at the natural-language level in practice.

As all types in \type are in the natural-language domain,
we leverage Large Language Models (LLMs) as the backend to generate and reason.
LLMs have demonstrated remarkable capabilities in understanding both code and natural language,
and with their reasoning abilities, can analyze program semantics effectively.

We implement \sys,
an extension of pytype~\cite{pytype} that incorporates \type reasoning for enhanced vulnerability detection in Python programs.
Following the code-execution flow inferred by pytype,
\sys maintains a dynamic representation of the program's state and prompts an LLM to update the \type of variables and expressions following the encountered operations (e.g., function calls or built-in operations).
We focus on detecting common vulnerabilities in web applications,
such as path traversal and SQL injection,
with security policies derived from Common Weakness Enumeration (CWE)~\cite{cwe}.

We evaluate \sys through comprehensive experiments on real-world Python codebases.
As a static code analyzer,
\sys should follow generic criteria for static analysis---
reporting as many genuine bugs as possible
while minimizing false alarms.
On a dataset of known vulnerabilities and their patches
(SVEN~\cite{sven-llm}),
\sys achieves \detectrate detection precision and \acrate accuracy,
demonstrating its effectiveness in vulnerability detection.
To address memorization concerns,
we evaluate \sys on recent CVEs
reported after the backend LLM's training cutoff,
and observe consistent performance across both datasets.

We also deploy \sys to analyze \nprojects popular open-source Python projects.
\sys discovered 15 
new vulnerabilities across high-profile codebases,
including an official Microsoft library,
Python Software Foundation (PSF) server infrastructure, and
systems maintained by the Internal Revenue Service (IRS).
Nine of these vulnerabilities have been confirmed
by respective development teams.
\sys demonstrates practical scalability in this process---%
analysis on the largest codebase (195 API endpoints and 26k Python bytecodes) was
completed in around fourteen hours with an estimated LLM inference cost of \$3.19 using equivalent {Qwen-Plus API pricing}\footnote{\url{https://www.alibabacloud.com/help/en/model-studio/models}},
making it a viable tool for real-world deployment. 

In summary, we make the following contributions:
\begin{itemize}[itemsep=1pt, topsep=2pt]
    \item A semantics-aware approach for vulnerability detection by introducing
        \type, a type system that enables vulnerability detection via automated code reasoning
        in natural language and harvests semantic information from source code that is not
        encoded in syntactic data structures (\autoref{s:design}).

    \item \sys, a prototype implementation that leverages LLMs to perform \type
        reasoning on Python programs to detect vulnerabilities common in web applications
        (\autoref{s:implementation}).

    \item Extensive evaluation demonstrating \protect\sys's effectiveness in detecting
        known vulnerabilities with high accuracy and precision, and its ability to discover
        new security issues in popular open-source projects while exhibiting practical
        scalability (\autoref{s:evaluation}).
\end{itemize}

\section{Background}
\label{s:background}

In this section,
we present key background information that motivates
the design of \sys, including
challenges in applying conventional static analysis
and similarity between type checking and vulnerability modeling. 

\subsection{Static Analysis for Vulnerability Detection}
\label{ss:static_analysis}

Static analysis has always been a crucial tool for
vulnerability scanning in software development~\cite{beller2016analyzing,ayewah2010google,distefano2019scaling,sadowski2018lessons,zhou2022non}.
While technically any tool that examines a program
without executing it with concrete inputs can be considered a
static analyzer~\cite{ayewah2008using},
in practice, most static analyzers
(e.g., CodeQL~\cite{codeql}, Bandit~\cite{bandit}, and semgrep~\cite{semgrep})
relies on predefined bug patterns expressed over syntactic features
such as AST, CFG, and PDG, which also include data flow information.

However, defining bug patterns is quite challenging.
To demonstrate, 
consider the path traversal vulnerability.
In a simplified scenario presented below, which both CodeQL and Bandit can detect,
an API endpoint \cc{get\_file} takes a \cc{file\_name} parameter from user input.
Without proper validation,
an attacker could inject directory traversal sequences
(e.g., \cc{../../etc/passwd}) into this parameter.
When combined with a base directory path (e.g., \cc{FILE\_DIR}),
this malicious input could allow access to files outside the intended directory scope,
potentially exposing sensitive files like \cc{/etc/passwd}.

\vspace{3pt}
\hfill\begin{minipage}[c]{\linewidth}\vspace{0.5em}\begin{Verbatim}[commandchars=\\\{\},numbers=left,firstnumber=1,stepnumber=1,codes={\catcode`\$=3\catcode`\^=7\catcode`\_=8\relax}]
\PY{n+nd}{@app}\PY{o}{.}\PY{n}{route}\PY{p}{(}\PY{l+s+s2}{\PYZdq{}}\PY{l+s+s2}{/get\PYZus{}file}\PY{l+s+s2}{\PYZdq{}}\PY{p}{,} \PY{n}{methods}\PY{o}{=}\PY{p}{[}\PY{l+s+s2}{\PYZdq{}}\PY{l+s+s2}{POST}\PY{l+s+s2}{\PYZdq{}}\PY{p}{]}\PY{p}{)}
\PY{k}{def} \PY{n+nf}{get\PYZus{}file}\PY{p}{(}\PY{p}{)}\PY{p}{:}
    \PY{n}{file\PYZus{}name} \PY{o}{=} \PY{n}{request}\PY{o}{.}\PY{n}{form}\PY{o}{.}\PY{n}{get}\PY{p}{(}\PY{l+s+s2}{\PYZdq{}}\PY{l+s+s2}{file\PYZus{}name}\PY{l+s+s2}{\PYZdq{}}\PY{p}{)}
    \PY{n}{file\PYZus{}dir} \PY{o}{=} \PY{n}{os}\PY{o}{.}\PY{n}{path}\PY{o}{.}\PY{n}{abspath}\PY{p}{(}\PY{n}{FILE\PYZus{}DIR}\PY{p}{)}
    \PY{n}{path} \PY{o}{=} \PY{n}{os}\PY{o}{.}\PY{n}{path}\PY{o}{.}\PY{n}{join}\PY{p}{(}\PY{n}{file\PYZus{}dir}\PY{p}{,} \PY{n}{file\PYZus{}name}\PY{p}{)}

    \PY{k}{with} \PY{n+nb}{open}\PY{p}{(}\PY{n}{path}\PY{p}{,} \PY{l+s+s2}{\PYZdq{}}\PY{l+s+s2}{rb}\PY{l+s+s2}{\PYZdq{}}\PY{p}{)} \PY{k}{as} \PY{n}{f}\PY{p}{:}
        \PY{n}{data} \PY{o}{=} \PY{n}{f}\PY{o}{.}\PY{n}{read}\PY{p}{(}\PY{p}{)}
    \PY{k}{return} \PY{n}{data}
\end{Verbatim}\end{minipage}\vspace{0.5em}
\vspace{3pt}

In this scenario,
the root cause is the direct use of user-controlled input in file path construction without proper validation or sanitization.
Detecting such vulnerabilities followingly requires identifying the data flow from user input to sensitive operations (e.g., file access) and recognizing the absence of necessary validation steps.
While this may seem straightforward, detecting such vulnerabilities effectively using static analysis is challenging and requires addressing two key objectives:

\textbf{1) A comprehensive knowledge database.}
The tool must understand the purpose and behavior of each function used in the code (e.g., \cc{os.path.join}, \cc{os.path.abspath}, and \cc{open}) and comprehend how they interact to form file paths. This requires maintaining a comprehensive knowledge base of functions and their security implications, which becomes increasingly challenging as programming languages and libraries evolve rapidly.
For instance, consider the \cc{pymupdf4llm.to\_markdown} function (shown in~\autoref{code:semantics-bug-exp}). 
Despite this library being installed millions of times per month\footnote{\url{https://pypistats.org/packages/pymupdf4llm}}, and its function signature clearly indicating that it reads a file, neither CodeQL nor Bandit recognizes this behavior, resulting in missed vulnerabilities.

\textbf{2) Detailed and versatile vulnerability patterns.}
The tool must recognize various sanitization and validation techniques that prevent vulnerabilities.
For instance, the patch for CVE-2025-6776 (shown below) adds input validation \cc{allowed\_file} in line 4, which CodeQL fail to recognize as a mitigation, although this behavior is semantically obvious:

\vspace{3pt}
\hfill\begin{minipage}[c]{\linewidth}\vspace{0.5em}\begin{Verbatim}[commandchars=\\\{\},numbers=left,firstnumber=1,stepnumber=1,codes={\catcode`\$=3\catcode`\^=7\catcode`\_=8\relax}]
\PY{n+nd}{@api}\PY{o}{.}\PY{n}{route}\PY{p}{(}\PY{l+s+s1}{\PYZsq{}}\PY{l+s+s1}{/upload\PYZus{}to\PYZus{}local}\PY{l+s+s1}{\PYZsq{}}\PY{p}{)}
\PY{k}{def} \PY{n+nf}{upload}\PY{p}{(}\PY{p}{)}\PY{p}{:}
    \PY{n}{image} \PY{o}{=} \PY{n}{request}\PY{o}{.}\PY{n}{files}\PY{o}{.}\PY{n}{get}\PY{p}{(}\PY{l+s+s1}{\PYZsq{}}\PY{l+s+s1}{image}\PY{l+s+s1}{\PYZsq{}}\PY{p}{,} \PY{k+kc}{None}\PY{p}{)}
    \PY{k}{if} \PY{n}{image} \PY{o+ow}{and} \PY{n}{allowed\PYZus{}file}\PY{p}{(}\PY{n}{image}\PY{o}{.}\PY{n}{filename}\PY{p}{)}\PY{p}{:}
        \PY{n}{path} \PY{o}{=} \PY{n}{os}\PY{o}{.}\PY{n}{path}\PY{o}{.}\PY{n}{join}\PY{p}{(}\PY{n}{UPLOAD\PYZus{}DIR}\PY{p}{,} \PY{n}{image}\PY{o}{.}\PY{n}{filename}\PY{p}{)}
        \PY{n}{image}\PY{o}{.}\PY{n}{save}\PY{p}{(}\PY{n}{path}\PY{p}{)}
    \PY{k}{else}\PY{p}{:}
        \PY{k}{raise} \PY{n}{ParameterException}\PY{p}{(}\PY{p}{)}
    \PY{k}{return} \PY{n}{Success}\PY{p}{(}\PY{p}{)}
\end{Verbatim}
\end{minipage}\vspace{0.5em}
\vspace{3pt}

Without understanding both vulnerable patterns and their mitigations, the tool risks missing actual vulnerabilities while raising false alarms on secure code.

\subsection{Type System and Vulnerability Modeling}
\label{ss:type_system}

A type system is a set of rules that dictates
how to assign and check types to various constructs in a program,
such as variables, expressions, functions, and modules~\cite{pierce2002types}.
Type systems can also be bidirectional when rules are not fixed a priori~\cite{dunfield2021bidirectional}.
The primary purpose of a type system is to ensure that operations in a program are performed on compatible types, thereby preventing type errors and enhancing program reliability.
A static type system further
requires type inference and checking at compile-time,
and forcing a program that fails the static type check to execution
(if possible) is almost doomed to trigger unexpected behaviors.

For example, in the code snippet below, the function \cc{retrieve\_email} (line 1) expects a string \cc{name} as its second argument. When the function is called with an integer (line 5), a static type checker would flag this as a type error. Failing to catch this mismatch and executing this code would lead to a runtime \cc{TypeError} exception.

\vspace{3pt}
\hfill\begin{minipage}[c]{\linewidth}\vspace{0.5em}\begin{Verbatim}[commandchars=\\\{\},numbers=left,firstnumber=1,stepnumber=1,codes={\catcode`\$=3\catcode`\^=7\catcode`\_=8\relax}]
\PY{k}{def} \PY{n+nf}{retrieve\PYZus{}email}\PY{p}{(}\PY{n}{db}\PY{p}{:} \PY{n+nb}{dict}\PY{p}{,} \PY{n}{name}\PY{p}{:} \PY{n+nb}{str}\PY{p}{)} \PY{o}{\PYZhy{}}\PY{o}{\PYZgt{}} \PY{n+nb}{str}\PY{p}{:}
    \PY{k}{return} \PY{n}{db}\PY{o}{.}\PY{n}{get}\PY{p}{(}\PY{n}{name}\PY{p}{,} \PY{l+s+s2}{\PYZdq{}}\PY{l+s+s2}{Email not found}\PY{l+s+s2}{\PYZdq{}}\PY{p}{)}

    \PY{c+c1}{\PYZsh{} db = \PYZob{}\PYZdq{}alice\PYZdq{}: \PYZdq{}alice@example.com\PYZdq{}\PYZcb{}}
    \PY{n}{retrieve\PYZus{}email}\PY{p}{(}\PY{n}{db}\PY{p}{,} \PY{l+m+mi}{23}\PY{p}{)}
\end{Verbatim}

\end{minipage}\vspace{0.5em}
\vspace{3pt}

Similar to the syntactical \cc{TypeError},
a vulnerability can be viewed as a
program construct of type $X$ participating
in an operation that expects type $Y$.
However,
unlike conventional type systems that have formally defined
type inference rules
(e.g., we can infer that \cc{v} is of \cc{str} type
from the assignment \cc{v = "abc"}),
it is hard to infer that a variable is
a ``\cc{user input}'' type
purely based on how the AST or CFG looks like, especially given the diversity of ways
to obtain user inputs in real-world codebases.
This is partially the reason that many rules
for finding the path traversal vulnerabilities focus on better definition of 
sources (APIs that return untrusted user inputs) and
sinks (APIs that consume paths and cause path traversal) in static analysis~\cite{zhao2024leveraging,li_iris_2025}.

The analogy between vulnerabilities and types extends to type hierarchies and compositions. Just as complex types can be built from simpler ones (e.g., \cc{List<String>}), complex vulnerabilities often arise from the interaction of multiple simpler conditions. 
For example, taking user input may not always lead to a vulnerability, and a file operation is not inherently vulnerable by itself. It is when these two conditions interact together in a specific way that can lead to a dangerous path traversal vulnerability.

\section{Towards LLM-assisted Static Analysis}    
\label{s:llm_analysis}

While it is intuitive to see why LLMs are
suitable candidates for addressing
certain limitations of static analysis
(e.g., finding sources and sinks autonomously, see~\autoref{s:background}),
how to systematically integrate LLMs into a vulnerability detection pipeline remains an open research challenge.
Here, we present
generic criteria for evaluating
code analysis tools,
followed by design options that we considered and justifications for our design choices.

\vspace{2pt}
\noindent
\textbf{Design goals} for an LLM-assisted static code analyzer:
\begin{itemize}[noitemsep]
\item \textbf{Correctness}: Avoid flagging safe code as vulnerable
(i.e., high precision);

\item \textbf{Completeness}: Avoid flagging vulnerable code as safe
(i.e., high recall);

\item \textbf{Automation}: Operate with minimal human input
such as manual review or configuration;

\item \textbf{Scalability}:
Analyze large codebases
within a reasonable time frame and monetary cost.
\end{itemize}

\subsection{Design Candidates}

\textbf{Option 1}:
Directly feed the entire source code and
the description of a specific CWE to an LLM, and
ask the LLM to conclude
whether the code contains vulnerabilities.
This option essentially offloads the entire task to the LLMs,
which have complete control over how they explore the codebase
and decide if the codebase contains vulnerabilities.

However, while modern LLMs have increased their input size capacity (up to 32k for Qwen3~\cite{qwen3}), they still face limitations when processing large codebases or complex programs. Indeed, many models are initially trained on short text of code snippets~\cite{repoaudit2025,roziere2023code} and struggle to maintain context when analyzing longer pieces of code.
Furthermore, LLMs tend to prioritize some parts of the input over others, potentially overlooking critical sections of code that are essential for vulnerability detection. Prior works have mixed results when applying LLMs directly to large codebases, with some studies reporting that less attention is paid to the middle of the input~\cite{liu2023lost}, while others have found that LLMs indeed focus less on the end~\cite{sovrano2025large}. As a result, directly feeding the entire codebase to LLMs is not effective and
may lead to missed vulnerabilities or inaccurate assessments~\cite{jimenez2023swe}.

Alternatively, one may consider breaking down the codebase into smaller chunks and feeding them individually to the LLM. However, existing works have shown that the loss of the context and interdependencies between different parts of the code can hinder the LLM's ability to accurately interpret the code~\cite{steenhoek2024err,khare2025understanding}.
An open question remains on how to effectively slice the codebase while preserving the necessary context for accurate vulnerability detection.

\textbf{Option 2}:
Task LLMs to create rules suitable for static analyzers to execute
(e.g., CodeQL rules curated to the target codebase).
This option essentially delegates rule authoring to LLMs,
leveraging their abilities to analyze source code and
generate code-like rules
while preserving the traditional static-analysis workflow.

However,
enumerating comprehensive vulnerability rules is challenging~\cite{7985689}, and certain vulnerabilities cannot be easily captured by rigid logical rules~\cite{li_iris_2025}.
Moreover, LLMs are inherently stochastic~\cite{vaswani2017attention,xia2025beyond}, which can lead to hallucinations~\cite{ji2023survey,zhao2023survey,li2024enhancing}. 
A recent study reports that the F1 score of vulnerability detection using LLM-generated rules is 0.18~\cite{li_iris_2025}, highlighting practical limitations.
As a consequence, generated rules necessitate additional verification prior to deployment (e.g., via manually crafted examples~\cite{li2025automated}), which constrains scalability and automation.

\subsection{Seeking a Middle Ground}

\textbf{Option 3 (our design)}:
Grounding code exploration in a deterministic and algorithmic traversal, while also allowing LLMs to comprehend source code semantics and
improvise vulnerability checking during code exploration.
This option strikes a balance between Options~1 and~2: deterministic, algorithmic code exploration provides guaranteed coverage (reducing the chance of missed bugs), while LLMs remain free to reason beyond rigid, predefined static-analysis rules and adapt flexibly.

Inspired by how vulnerability modeling is similar to type systems
(\autoref{ss:type_system}),
we adopt bidirectional typing~\cite{dunfield2021bidirectional}
as algorithmic code exploration procedure
(which touches every reachable part of the program)
and allow LLMs to reason about vulnerabilities in
a localized type checking context.
By sending small queries (approximately 200 tokens),
we keep LLMs focus on only one task.
This also reduces uncertainty of LLM responses.
At the same time, we do not restrict LLMs' outputs to predefined templates, avoiding the brittleness of logical analysis rules by allowing LLMs to reason in natural language directly.

However, while this option preserves the strengths of static analysis and LLM reasoning, we acknowledge that
it also inherits certain limitations from both sides
(e.g., imprecise point-to results from static analysis and
extra cost and non-determinism from LLM queries).
More detailed evaluation and discussion on the limitations
can be found in~\autoref{ss:case-study}, \ref{ss:zero-day} and~\autoref{s:limitations}.

\section{Design}
\label{s:design}

In this section,
we describe the design details of \sys---%
an \type-aided static analyzer for Python programs.

\begin{figure*}[t]
    \centering \includegraphics[width=0.99\textwidth]{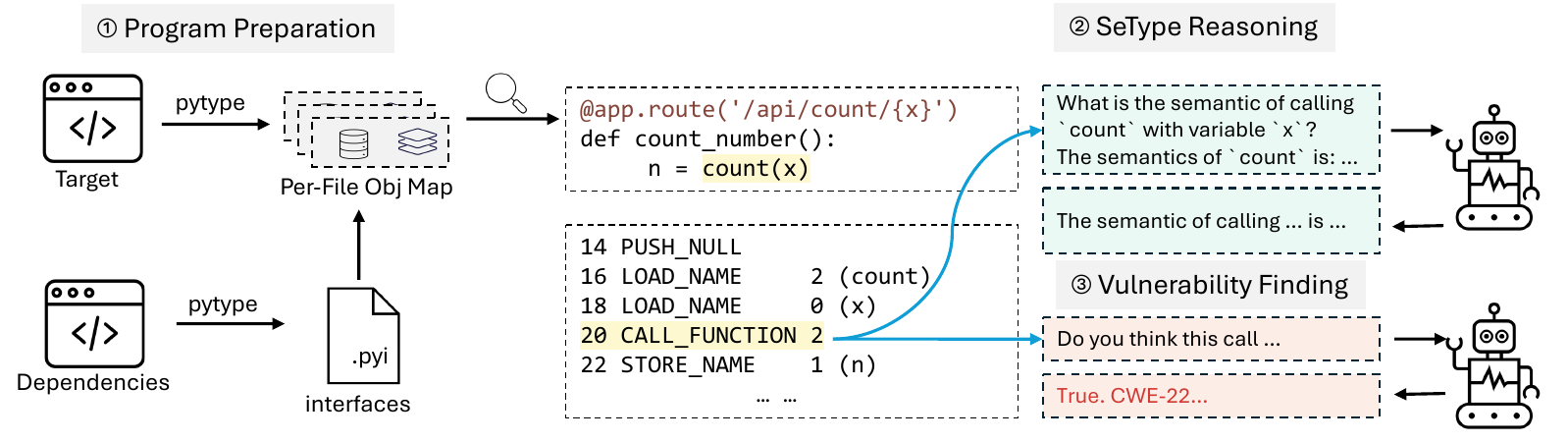}
      \caption{Pipeline of \sys. \sys starts with preparing the program through pytype by generating interface stubs and program maps, and identifying the entry points (\protect\WC{1}). 
      Next, it simulates the program execution by following the program's IPCFG starting from the identified endpoints, during which it interacts with an LLM to infer the \type of variables (\protect\WC{2}) and perform type checking to identify potential vulnerabilities (\protect\WC{3}).}
  \label{fig:pipeline} 
\end{figure*}

\PP{Overview}
\sys is a static-analysis framework that identifies potential vulnerabilities in Python programs using \type.
The analysis process traverses the program's CFG while performing \type checking that combines syntactic type information with semantic reasoning about variable usage, derived from non-executable context such as variable names, comments, and docstrings.
We illustrate this process with an example in~\autoref{fig:pipeline}.

\WC{1} \sys begins by running vanilla pytype on the project to generate interface stubs for dependencies and build the program's CFG. This preprocessing step produces per-file object maps containing syntactic type information and control flow structures, which will be used in subsequent LLM-assisted analysis.

Using the inferred syntactic types,
\sys identifies entry points through the \cc{main} function and common server endpoints (e.g., \cc{@app.route()} in Flask and \cc{re\_path()} in Django), then simulates program execution from these points by traversing the CFG at the bytecode level via an extended pytype virtual machine. Entry points are analyzed once, while other functions are revisited on each call to capture execution context precisely.

\WC{2} During the simulation, \sys initializes the \type of variables based on the non-executable context and instruments the bytecodes to maintain the \type of variables and expressions encountered.
For each operation that may change the \type (e.g., assignment, function call, branching, and loop), \sys queries the LLM for the updated \type after the operation with a prompt that includes the current \type of the involved variables and the non-executable context. The response from the LLM is then parsed and stored as the updated \type of the target variable or expression.
Example prompts and the involved \type before and after the operation are illustrated in \autoref{fig:walkthrough-bug}.

\WC{3}
\sys uses \type to identify potential vulnerabilities.
For each operation in the program, \sys follows a set of predefined vulnerability rules to determine whether it may introduce a security risk. If a mismatch is detected between the \type of the involved variables and the expected \type defined in the vulnerability rule, \sys flags it as a potential vulnerability.

\subsection{Program Preparation}
\label{ss:preparation}
\sys builds upon pytype, a static type analyzer for Python that simulates bytecode execution to infer and check syntactic types~\cite{pytype}. 
Rather than reimplementing static analysis from scratch, \sys extends pytype's infrastructure by: (1) leveraging its CFG, bytecode-level virtual machine, and per-file object maps for syntactic type information; (2) augmenting the \cc{variable} and \cc{frame} abstractions to track \type and execution contexts; and (3) instrumenting the VM to query an LLM at semantically important instructions.
This design enables \sys to combine pytype's robust static analysis with LLM-assisted semantic reasoning.

\PP{Preprocessing} \sys first runs vanilla pytype on the project's dependencies to generate interface stubs. These stubs capture the syntactic interfaces of external libraries and allow subsequent analysis to resolve types and call signatures for third-party and standard libraries. As LLMs are already trained on common public libraries, providing accurate stubs for dependencies is typically sufficient for LLMs to reason about the \type of variables involving these libraries.

After generating dependency stubs, \sys runs pytype on the target project to produce per-file object maps and to build the program's CFG. pytype records syntactic types of variables at the basic-block level. These artifacts are then augmented with \type and execution-context during the LLM-assisted analysis stage.

\PP{Endpoint Identification} \sys first checks the \cc{main} function, which is the common entry point for Python scripts. 
Additionally, \sys detects common endpoints to web frameworks
by matching framework-specific routing patterns.
This is straightforward as we can confirm
which Python objects are framework-related
by examining the syntactic types of these objects inferred by pytype. 

For Flask, \sys treats functions decorated with \cc{app.route(...)} or \cc{bp.route(...)} as endpoints, as long as the container object (e.g., \cc{app} or \cc{bp}) is recognized as \cc{flask.app.Flask} or \cc{flask.blueprints.Blueprint} by pytype. A similar approach is applied to FastAPI. 
For Django, \sys extracts endpoints referenced in URL configuration (e.g., \cc{path(...)} and \cc{re\_path(...)}) and treats the referenced callables as endpoints. \sys additionally supports class-based views by recognizing view classes via pytype's type information: when pytype identifies a class inheriting from Django's view base classes (e.g., \cc{django.views.View}), \sys treats typical handler methods (such as \cc{get}, \cc{post}, and \cc{dispatch}) defined in the class as endpoints.
All identified endpoints serve as starting points for the LLM-assisted \type inference (\WC{2}) and
vulnerability finding process (\WC{3}).

\subsection{\type Inference}
\label{ss:type-inference}

\begin{table*}[t]
    \centering
    \small
    \caption{Sample prompt templates for \type tracking in different operations. Comments and docstrings are included in the actual prompts but omitted here for brevity. Texts in \cc{monospace} are dynamically filled based on the analysis context.}
    \label{tab:prompts}
    \begin{tabular}{lp{.85\textwidth}}
    \toprule
    \textbf{Operation} & {\textbf{Prompt Template}}  \\\midrule
    Assignment & I store the semantic \cc{rhs\_type} to a variable named \cc{lhs\_name}. Please describe the semantic of this new variable.\\
    Call & I am calling a function named \cc{func\_name}. This call takes the following parameters: \cc{param\_list}. The semantic of the parameters are as following: \cc{param\_types}. Please define the semantic of the output of this function call.\\
    Merge & After statically analyzing all possible executing branches, one variable is possible to have these possible semantic in different branches: \cc{semantic}. Please consider all these possibilities and summarize its correct semantic.\\
    Loop & Variable \cc{var\_name} has semantics \cc{old\_type}. Now, I put this variable into a loop through an iterator \cc{iter}. After the first iteration, the semantics of \cc{var\_name} becomes into \cc{new\_type}. Please define the final semantics of this variable after all iterations.\\
    Branching & I compare whether the lhs variable is \cc{operator} the rhs variable. The lhs variable has the semantic \cc{lhs\_type}. The rhs variable has the semantic \cc{rhs\_type}. Suppose the comparison result is \cc{condition}, summarize how this \cc{condition} influences the execution of following code. \\
    \bottomrule
    \end{tabular}
\end{table*}

We formally define \type in~\autoref{app:type-system}. Here we describe how \sys infers \type during program analysis.

For program $P$, we extract its CFG $\mathbb{G} = (\mathbb{B}, \mathbb{E})$, where $\mathbb{B}$ is the set of basic blocks and $\mathbb{E}$ is the set of edges connecting the blocks. 
The analysis process traverses $\mathbb{G}$ and starts from the entry points identified in \autoref{ss:preparation}. \type of all variables are initialized to empty.

\sys analyzes each basic block in topological order within functions and depth-first order across functions. 
For each expression, \sys updates the variable's \type by querying the LLM using templates from \autoref{tab:prompts}. 
For example, in an assignment operation, the left-hand side variable's \type is updated based on the right-hand side expression's \type and the name of the left-hand side variable.
If the right-hand side is a function call, \sys steps into the callee and infers the output's \type before updating the left-hand side variable's \type.
These \type are bounded to the corresponding basic block, and are inherited to successor blocks following the CFG if it is the only predecessor.
Alternatively, if a basic block has multiple predecessors (e.g., due to branching or loops), \sys merges the \type from all predecessors by querying the LLM with the merge prompt template in \autoref{tab:prompts}.

To handle variable updates within loops, \sys employs an induction process that considers three key aspects: (1) the variable's \type prior to entering the loop, (2) the variable's \type after completing one iteration of the loop, and (3) the \type of the loop iterator. By querying the LLM with these three \type, \sys infers the variable's \type upon exiting the loop.

Beyond variable \type, \sys also captures the execution context of each basic block.
For example, as illustrated below, a branching statement checks a variable's type before an operation, 
the execution context within that branch thus contains additional information about the variable's \type. This contextual information is essential for accurate \type analysis and vulnerability detection. 
\hfill\begin{minipage}[c]{\linewidth}\vspace{0.5em}\begin{Verbatim}[commandchars=\\\{\},numbers=left,firstnumber=1,stepnumber=1,codes={\catcode`\$=3\catcode`\^=7\catcode`\_=8\relax}]
\PY{k}{if} \PY{n}{IS\PYZus{}ADMIN}\PY{p}{:}
    \PY{n}{user} \PY{o}{=} \PY{n}{admin\PYZus{}user}\PY{p}{(}\PY{p}{)}
\end{Verbatim}
\end{minipage}\vspace{0.5em}

For each basic block, we define its execution context as a set of conditions that must be satisfied for the block to be executed. These conditions are derived from the branching statements leading to the block in the CFG. We use the ``branching'' templates presented in~\autoref{tab:prompts} to capture the execution context during the analysis.

\subsection{Vulnerability Finding}
\label{ss:bug-finding}

With the inferred \type and execution contexts,
\sys performs type checking to identify potential vulnerabilities in the program.
Each time a function that matches a vulnerability rule is called
(see details below on how \sys checks for matches),
\sys checks whether
the \type of the arguments passed into the operation satisfy the
expected \type of the operation defined in the vulnerability rule.
If a type mismatch is detected, \type flags it as a potential vulnerability.

\PP{Vulnerability Rules}
We focus on common, high-impact vulnerabilities, including SQL injection, command injection, cross-site scripting (XSS), and path traversal (see~\autoref{app:taret-cwes} for the complete list of target CWEs). 
Each vulnerability rule is described following CWE definitions~\cite{cwe} and maintains a high-level abstraction to facilitate generalization across different codebases.

To mitigate limitations of LLM reasoning, \sys employs a checklist-based verification strategy~\cite{cook2024ticking}. The LLM is queried about each item independently, and \sys aggregates the results to decide whether a potential semantic violation exists.
For example, the path traversal vulnerability is defined with the following checks:

\begin{itemize}[noitemsep,topsep=1pt]
    \item is accessing (e.g., reading, writing, or deleting) a path;
    \item has the accessed path constructed with a value provided by an external client;
    \item and, the called function is not a safe access function that automatically sanitize the accessed path.
\end{itemize}

\PP{Capturing Mismatches}
\sys limits vulnerability checks to function calls whose callees are not defined within the analyzed project (i.e., external library functions). 
For each external function call, \sys constructs a prompt containing the current \type of involved variables and queries the LLM to match against vulnerability rules. Upon finding a match, \sys queries the LLM again with the execution context to confirm the rule's conditions are satisfied. If confirmed, the operation is flagged as a potential vulnerability; otherwise, analysis continues without raising an alert.

To reduce LLM queries, \sys employs taint analysis to track untrusted data flows from external sources. If data involved in an operation is identified as tainted, \sys skips the LLM query checking whether the data originates from an external client.

\subsection{Illustrative Example}
\label{ss:walkthrough}

\begin{figure*}[t]
    \centering
    \centering \includegraphics[width=0.98\textwidth]{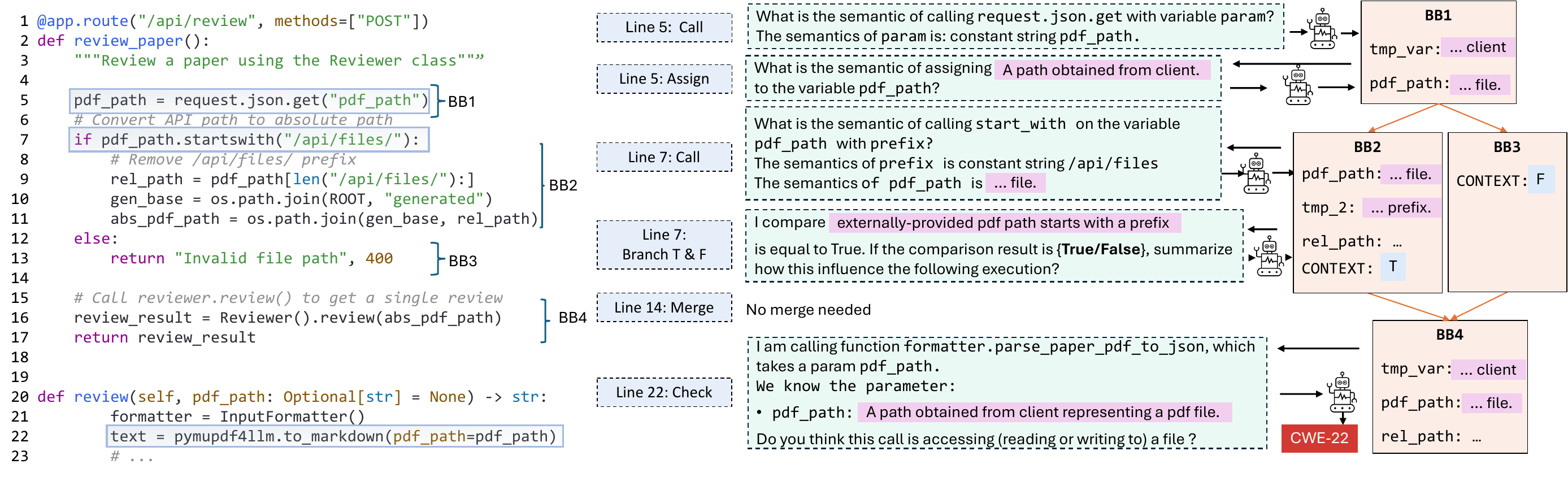}
    \caption{Walkthrough with a simplified path traversal vulnerability adopted from CVE-2025-55149~\cite{tinyscientist}. The code snippet is the same as the one presented in~\autoref{code:semantics-bug-exp}. We illustrate how \sys simulates the program execution following the CFG, infers the \type of variables, captures the execution context, and identifies the potential security vulnerability. We only illustrate the key communications with the LLM for brevity and clarity.}
    \label{fig:walkthrough-bug}
\end{figure*}

Putting all the components together,
we present an illustrative example in~\autoref{fig:walkthrough-bug}, which is a simplified version of a vulnerability adopted from CVE-2025-55149~\cite{tinyscientist}, same as the one shown in~\autoref{code:semantics-bug-exp}. In this example, we skip the program preparation step (\WC{1} in~\autoref{fig:pipeline}) and start with \type analysis (\WC{2})
and vulnerability finding (\WC{3}).

\sys starts simulating the program execution from the program entry point, which is the \cc{review\_paper} function in this example. 
The CFG of this function contains four basic blocks: the entry block (BB1), the \cc{if} block (BB2), the \cc{else} block (BB3), and the exit block (BB4). Analysis starts from the entry block (BB1). 
In line 5, \sys encounters a function call \cc{request.json.get("pdf\_path")} that retrieves a file path from the incoming request. 
\sys constructs a prompt using the call template in \autoref{tab:prompts} and queries the LLM to infer the \type of the call output. 
Given that this output is not assigned to any variable, \sys temporarely stores it in a special variable \cc{tmp\_var} with the inferred \type ``a file path provided by an external client''. 
Still in this line, an \cc{assign} operation assigns the value of \cc{tmp\_var} to the variable \cc{pdf\_path}. 
\sys constructs another prompt using the assignment template in \autoref{tab:prompts} and queries the LLM to update the \type of \cc{pdf\_path} to ``a path provided by an external client pointing to a pdf file''.
Both types are stored in the \cc{BB1} object in the pytype virtual machine.

With some executions in between, the analysis reaches line 7, where \sys encounters a branching statement that checks whether the file at \cc{pdf\_path} has a specific prefix. \sys constructs a prompt using the branching template in \autoref{tab:prompts} and queries the LLM to capture the execution context for both branches. The resulting execution contexts are stored in \cc{BB2} and \cc{BB3} of the pytype virtual machine, respectively. Since these two basic blocks inherit from \cc{BB1}, the \type of \cc{pdf\_path} is also inherited.

After analyzing both branches, the analysis reaches line 14, where the two branches merge. \sys similarly merges the execution contexts of \cc{BB2} and \cc{BB3}. However, since no variable is assigned in either branch, the \type of all variables remains unchanged.

Finally, when the analysis reaches line 21, \sys encounters a function call to \cc{pymupdf4llm.to\_markdown} that processes the file at \cc{pdf\_path}. \sys queries the LLM to check whether this operation is vulnerable. The LLM confirms that this operation involves path access, and the accessed path is constructed from untrusted input and is not sanitized, matching all conditions in the path traversal rule. 
Consequently, \sys flags this operation as a potential vulnerability.

\section{Implementation}
\label{s:implementation}

We implement \sys, an extension of pytype that incorporates LLM-based \type reasoning into syntactic type analysis for Python projects. The implementation extends pytype release 2024.10.11~\cite{pytype} with approximately 3,000 lines of code to support whole-project semantic analysis and Python 3.10. We use the Qwen3 32B model~\cite{qwen3} as the LLM backend via the Ollama API~\cite{ollama}, with \cc{top\_p=0.3} to reduce output randomness. All other parameters are at default values.

\sys targets three popular Python web frameworks: Flask, FastAPI, and Django. 
We implement framework-specific rules to identify application endpoints, as detailed in~\autoref{ss:preparation}.
The system focuses on \nvul common web vulnerability types, including SQL injection, XSS, and command injection. The complete list of target vulnerabilities is provided in~\autoref{app:taret-cwes}. 
For each vulnerability type, we design tailored prompts to assist the LLM in detecting \type inconsistencies associated with that vulnerability, adhering to the principles outlined in~\autoref{ss:bug-finding} and employing the zero-shot prompting method.

\section{Evaluation}
\label{s:evaluation}

In this section,
we evaluate \sys against the design goals
in \autoref{s:llm_analysis}
using real-world vulnerabilities. 

\subsection{Evaluation Setup}
\label{ss:setup}
  
\PP{Environments}
We run \sys on a server with
AMD EPYC 7302 CPU and
NVIDIA A100 40GB GPU.

\PP{Comparison Baselines} We compare \sys with three baseline tools:

\noindent
\textbullet\enspace\emph{CodeQL}~\cite{codeql} is a widely used \emph{static-analysis} platform that detects security issues via built-in and custom queries. It supports multiple programming languages, including Python. We use CodeQL release 2.22.3 and the default Python security queries in our evaluation. 
  
\noindent  
\textbullet\enspace\emph{Bandit}~\cite{bandit} is a Python-specific \emph{static-analysis} tool developed by the Python Code Quality Authority (PyCQA). We run Bandit with its default rule set. 

\noindent
\textbullet\enspace\emph{Vulnhuntr}~\cite{vulnhuntr} is an LLM-based vulnerability detection tool developed by Protect AI (Palo Alto Networks). It \emph{combines static analysis with LLM reasoning}: the LLM first inspects the codebase and highlights snippets it considers potentially vulnerable, along with a list of functions or variables it wants to examine further. 
Vulnhuntr then uses static analysis to locate the definitions of these symbols and feeds the gathered snippets back to the LLM. 
This iterative process continues until the model requests no additional context. 
We use Vulnhuntr's default configuration, which asks the LLM to assign a confidence score (0-10) and reports findings with score $\geq$7. 
We also patch Vulnhuntr to resolve certain parsing issues when it interacts with the LLM by extending its parser. Specifically,
we extend Vulnhuntr's parser to be more tolerant of common formatting deviations and to retry up to three times upon parsing failures.

\PP{Datasets}
We use three datasets to evaluate the effectiveness of \sys in detecting real-world vulnerabilities. Details of the construction and filteration of the three datasets is provided in ~\autoref{app:popular-projects}.

\noindent
\textbullet\enspace\emph{SVEN: Real-world Vulnerable Programs with Patches.}
We use the SVEN dataset~\cite{sven-llm}, which contains real-world vulnerable Python programs with corresponding developer patches.
After filtering out C projects, Python 2 projects, and those incompatible with current framework versions, we obtain 80 vulnerability instances from 27 repositories.
The dataset spans four CWE classes: path traversal, OS command injection, XSS, and SQL injection, with vulnerabilities patched between 2016 and 2018.
We evaluate on 160 total program versions (vulnerable and patched pairs), enabling comprehensive assessment of both detection recall and patch-aware accuracy across all instances.

\noindent
\textbullet\enspace\emph{Recent CVEs with (Partially Available) Patches.} 
To address the potential risk of SVEN containing outdated examples present in the LLM training data, we constructed a more recent CVE-based dataset. 
We queried the CVE database~\cite{cve-org} for vulnerabilities in projects utilizing Flask, Django, or FastAPI, focusing on CVEs published between 2024 and September 2025 (after the LLM's knowledge cutoff of end-2023
\footnote{\url{https://github.com/QwenLM/Qwen3/issues/525\#issuecomment-2159944330}}).

After filtering out entries that are not server applications implemented with one of the three frameworks or whose vulnerabilities do not originate from the Python component, we identified 41 candidate CVE entries. We then excluded entries that (1) do not fall within any CWE category targeted by \sys or (2) are not open-sourced on GitHub. The final dataset comprises 14 program versions corresponding to 9 CVEs spanning five CWE classes, with 5 of them patched by the developers.

\noindent
\textbullet\enspace\emph{Popular Open-Source Projects.} We evaluate \sys on a curated set of popular Python web projects that use Flask, Django, or FastAPI. 
We began with GitHub dependency data and used ghtopdep~\cite{ghtopdep} to automate repository collection. 
From the resulting candidates, we inspected the top 100 starred repositories per framework and removed tutorials, plugins/metapackages, toy examples, and mismatched repositories (e.g., 
repositories that only mention a framework in lock files or tests). After manual validation, deduplication, and exclusion of misclassifications, the final evaluation set comprises \nprojects repositories.

\subsection{Vulnerability Detection}
\label{ss:effectiveness}

\PP{Set up}
We evaluate \sys on the SVEN dataset and compare it against the baselines described in~\autoref{ss:setup}.
We run all tools on the vulnerable versions to assess detection recall and on the patched versions to measure the detection accuracy. Following our methodology, we analyze entire projects rather than individual files to obtain realistic vulnerability detection results.
For each tool, especially Bandit and CodeQL that report numerous warnings (e.g., hardcoded password string), we manually inspect the findings reported by the tool to check if they are related to the original vulnerabilities. 
We report four standard metrics: precision, recall, accuracy, and F1 score in~\autoref{tab:eval_sven}.

\PP{Compare to Baselines}
We observe that \sys effectively detects real-world vulnerabilities, outperforming all baselines in all four metrics. 
While CodeQL achieves a similar high accuracy, its recall is significantly lower than \sys, indicating that many vulnerabilities are missed, aligning with the observations in prior work~\cite{li_iris_2025}.
Bandit achieves an overall high performance but is still outperformed by \sys in all metrics.
Most notably, Bandit achieves a relatively low accuracy, potentially due to its coarse-grained heuristics that lead to many false positives (e.g., flagging all use of Python string concatenation in SQL construction as vulnerable), which produces overly generalized findings.
Vulnhuntr shows the lowest performance among all tools, which may be due to its design of depending on the LLM to analyze large code sections at once, making it more prone to hallucination and incomplete analysis.

\begin{table}[t]
    \centering
    \caption{Performance evaluation on the SVEN dataset.}
    \label{tab:eval_sven}
    \begin{tabular}{l|cccc}
        \toprule
        Tool & Precision & Recall & Accuracy & F1 Score \\\midrule
        CodeQL & 0.67 & 0.39 & 0.83 & 0.49 \\
        Bandit & 0.75 & 0.84 & 0.70 & 0.79 \\
        Vulhuntr & 0.52 & 0.45 & 0.53 & 0.48 \\
        \sys & \textbf{0.87} & \textbf{0.87} & \textbf{0.88} & \textbf{0.87} \\\bottomrule
    \end{tabular}

\end{table}

\PP{Validation through Recent CVEs}
To validate that the effectiveness of \sys is not due to potential data leakage in the LLM training data, we further evaluate \sys on recent CVEs after the knowledge cutoff of the used LLM.
Using the recent CVE dataset constructed in \autoref{ss:setup}, 
we similarly run all the tools on the vulnerable and patched versions (if the vulnerable has been patched) and check the warnings. The results are shown in~\autoref{tab:eval_cves}.

All tools show a decrease in performance.
This is possibly because that the programs in this dataset are more complex and the vulnerabilities are more sophisticated.
Specifically, we see that the precision of the two static analysis tools, CodeQL and Bandit, keep consistent compared to the results in~\autoref{tab:eval_sven}, while their accuracy and recall drop significantly. This indicates that while the static analysis rules are still effective in identifying certain vulnerabilities, they miss (potentially new) vulnerability patterns that are not covered by the rules.
The performance of Vulnhuntr and \sys drop slightly, indicating that the LLM unlikely has seen these vulnerabilities in its training data, and the detection capability of LLMs generalizes to unseen vulnerabilities.

\begin{table}[t]
    \centering
    \caption{Performance evaluation with recent CVEs.}
    \label{tab:eval_cves}
    \begin{tabular}{l|cccc}
        \toprule
        Tool & Precision & Recall & Accuracy & F1 Score \\\midrule
        CodeQL & 0.67 & 0.22 & 0.43 & 0.33 \\
        Bandit & 0.75 & 0.33 & 0.50 & 0.46 \\
        Vulhuntr & 0.67 & 0.44 & 0.50 & 0.53 \\
        \sys & \textbf{0.78} & \textbf{0.78} & \textbf{0.71} & \textbf{0.78} \\\bottomrule
    \end{tabular}
\end{table}

\subsection{Execution Cost}
\label{ss:cost-analysis}

\begin{table}[t]
    \centering
    \caption{Average execution time ($t$) and monetary cost ($c$) on the SVEN and new CVE dataset (in seconds and US cents). $t_p$ of \sys refers to the time for static analysis and type inference (\protect\WC{1} in~\autoref{fig:pipeline}).}
    \label{tab:eval_time}
    \begin{tabular}{l|cc}
        \toprule
        Tool (Metric) & SVEN & CVEs  \\\midrule
        CodeQL ($t$) & 14.92 & 16.60 \\
        Bandit ($t$) & 6.23 & 9.66 \\
        Vulhuntr ($t$) & 230.90 & 413.64 \\
        \sys ($t_p$) & 48.59 & 200.61 \\
        \sys ($t$) & 1470.52 & 3558.59 \\ \midrule
        Vulhuntr ($c$) & 2.16 & 3.56 \\
        \sys ($c$) & 10.28 & 26.44 \\\bottomrule
    \end{tabular}
\end{table}

We present the average of execution time and monetary cost of running all four tools on both the SVEN and recent CVE datasets in~\autoref{tab:eval_time}, following the \emph{scalability} design goal in~\autoref{s:llm_analysis}.
For monetary costs,
as we deploy the LLM locally,
we estimate the cost should a remote LLM service be used
based on tokens sent to and received from the LLM,
using Qwen-Plus pricing
(0.4 USD per 1M input tokens, 1.2 USD per 1M output tokens)
as a reference since it offers similar performance to our locally deployed Qwen3-32b model.

Two traditional static tools (CodeQL and Bandit) complete the analysis within seconds on average. Comparably, Vulnhuntr and \sys take significantly longer time due to the multiple rounds of LLM invocations. 
Notably, running the perparation phase of \sys (\WC{1} in~\autoref{fig:pipeline}) alone is already slower than CodeQL and Bandit, as it involves static type inference for all bytecode instructions in the project.
Compared with Vulnhuntr, \sys takes a longer time due to its more systematic and in-depth analysis of each endpoint, which involves multiple rounds of LLM queries per endpoint. 
Between the two datasets, analyzing the CVE dataset takes longer time for all tools, validating the observation that the CVE dataset contains more complex programs and vulnerabilities.

\subsection{Case Studies}
\label{ss:case-study}

While in the previous sections we have shown the overall effectiveness of \sys in vulnerability detection, we hereby present certain cases to help understand the strengths and weaknesses of \sys. Beyond the two discussions below, we also observe that \sys raises a false alarm due to an insufficient patch by the developers. We discuss this case in ~\autoref{app:insufficient-patches}.
\subsubsection{Cases Where \sys is More Effective}

We present two concrete cases where \sys outperforms all baselines: the example illustrated in~\autoref{fig:walkthrough-bug} and CVE-2025-49126~\cite{CVE_2025_49126}, in which user input is directly returned by an endpoint and used to generate HTML via \cc{get\_swagger\_ui\_html}, a FastAPI function that produces an HTML page. This results in an XSS vulnerability if the user input is maliciously crafted.

\PP{Compare with Traditional Static Analysis}
\sys leverages LLMs to reason about the security implications of functions without the need for predefined rules. 
In both examples, the vulnerabilities are caused by the use of (relatively) uncommon functions (i.e., \cc{get\_swagger\_ui\_html} and \cc{pymupdf4llm.to\_markdown}) which are not marked as dangerous sinks by static analysis tools.
As a result, CodeQL and Bandit fail to detect these vulnerabilities, while \sys successfully detects them using the LLM's knowledge and inference capabilities.

\PP{Compare with LLM-based Analysis}
\sys adopts a checklist-based analysis to guide the LLM to perform systematic reasoning, to limit the analysis complexity, and to follow execution paths to reduce hallucination.
We observe that \sys can understand the context of code snippets and reason effectively about their security implications.

In the example illustrated in~\autoref{fig:walkthrough-bug}, 
we observe that \sys precisely identifies the root cause: ``while there are checks for path prefixes and file extensions, and some validation of path components, there is \emph{no explicit check that normalizes the path} or ensures that the final resolved path does not escape the intended directory structure.''

In contrast, Vulnhuntr's approach of asking the LLM to analyze large code sections in a single pass makes it more prone to hallucination and incomplete analysis.
In its analysis of CVE-2025-49126, Vulnhuntr initially identifies two potentially vulnerable functions: one containing the vulnerability and the other benign.
However, when the LLM is asked for deeper analysis, it shifts focus to the benign function \emph{only} and later contradicts its initial assessment by concluding that both functions are unlikely to be vulnerable.
In this case, the LLM's inability to maintain focus on the vulnerable function leads to a missed detection by Vulnhuntr.

In the other example, Vulnhuntr hallucinates a nonexistent \cc{serve\_files} function and reports a path traversal vulnerability in it with high confidence (9/10).
These false detections highlight the challenges of applying LLMs to large codebases without structured guidance, in contrast to \sys's checklist-driven approach, which focuses the LLM's reasoning on specific execution paths and security properties.

\subsubsection{Cases Where \sys Misdetects}
\label{sss:missing-detections}

Although \sys combines static analysis with LLM-guided reasoning to balance their complementary strengths, limitations in either component can cause misdetections.

\PP{Statically Unreachable Code}
\sys works by iteratively analyzing the code snippets following the execution paths in the target project. It is unable to analyze code that is not reachable from the defined entry points.
For example, in the {SocialniSystem.cz}\footnote{\url{https://github.com/pirati-web/socialnisystem.cz}} project in the SVEN dataset, there is a vulnerability in the \cc{get\_context\_data} method.
This method overrides a parent class's method and is called by the Django framework through polymorphism, rather than being directly invoked in the main execution paths.

Another example is the use of dynamic features in Python, exemplified by the ScoringEngine\footnote{\url{https://github.com/DSU-DefSec/ScoringEngine}} project in the SVEN dataset.
It dynamically decides which module to import based on specifications in a configuration file, which is beyond the capability of static analysis.
These uncaught interactions make certain code snippets unreachable for \sys, and thus vulnerabilities in those snippets are missed by \sys.
We present an additional example in~\autoref{app:uncaught-interactions} to illustrate the complexity of such unreachable code.

\PP{LLM Hallucination}
LLM reasoning can be inconsistent, producing both false negatives and false positives. 
For example, in the Python Wiki App\footnote{\url{https://github.com/Pumala/python_wiki_app_redo}} there are five similar SQL-injection sites: \sys detects four but misses one because the LLM incorrectly judges \cc{db.query} as non-sink in that case while correctly identifying it as an execution sink elsewhere. 
In another case (CLA Hub\footnote{\url{https://github.com/stevetasticsteve/CLA_Hub}}), the model identifies the concrete template string (\cc{CE/home\_page.html}) as user-controlled and produces a false finding.

\subsection{Finding Zero-day Vulnerabilities}
\label{ss:zero-day}

\begin{table*}[t]
    \centering
        \caption{Average metrics of running \sys on the real-world popular projects. $t_p$ and $t$ refer to the time in the preparation phase only (\protect\WC{1} in~\autoref{fig:costs}) and the total execution time respectively, and are in seconds. $c$ is the LLM execution cost in US cents, \# API is the number of analyzed endpoints, \# Inst is the number of analyzed bytecodes, \# Inf is the number of LLM queries for inferring \type, and \# Sec is the number of LLM queries for checking security properties.}
        \label{tab:eval_real}
        \begin{tabular}{l|cccccc|cccccc}
            \toprule
             & \multicolumn{6}{c}{Per-Project} & \multicolumn{6}{c}{Per-API} \\\midrule
            Metric & $t_p$ & $t$ & $c$ & \# API & \# Inf & \# Sec & $t_p$ & $t$ & $c$ & \# Inst & \# Inf & \# Sec \\\midrule
            Flask & 168.08 & 1964.21 & 17.77 & 8.85 & 828.50 & 396.74 & 21.08 & 263.09 & 2.40 & 200.86 & 112.57 & 51.72 \\
            Django & 768.11 & 5221.27 & 29.60 & 24.86 & 1453.39 & 702.87 & 69.31 & 197.58 & 0.83 & 68.20 & 39.47 & 20.55 \\
            FastAPI & 297.78 & 2426.63 & 21.33 & 12.64 & 1179.88 & 446.70 & 33.96 & 210.50 & 1.73 & 174.03 & 94.32 & 36.44 \\\midrule
            All & 419.35 & 3250.75 & 23.03 & 15.66 & 1168.93 & 523.83 & 42.06 & 223.34 & 1.64 & 145.90 & 81.49 & 36.03 \\\bottomrule
             
        \end{tabular}
\end{table*}

To further demonstrate the effectiveness of \sys, we run it on popular open-source projects to discover zero-day vulnerabilities. 
We explore the top-starred Python projects on GitHub that use one of the three frameworks, as described in \autoref{ss:setup}.
We run \sys on the latest versions of these projects and manually inspect the findings reported by \sys to identify potential zero-day vulnerabilities. 

We identify 15 previously unknown vulnerabilities that we consider security-critical, and 2 additional vulnerabilities that are less severe due to a limited attack surface as the server is configured for local, personal use. The identified vulnerabilities were responsibly disclosed to the developers, and 9 have been confirmed by the developers at the time of writing. Details of the disclosure process are provided in the ethics section.

Of the 15 identified vulnerabilities, 9 are path traversal, 5 are open redirect, and 1 is SQL injection.
The categories of the identified vulnerabilities are consistent with those in the SVEN and recent CVEs datasets, indicating that these types of vulnerabilities remain prevalent in real-world applications.
However, we note that four out of five open-redirect vulnerabilities occur in projects using the Django framework, and eight out of nine path traversal vulnerabilities occur in projects using the Flask framework, suggesting that certain frameworks may be more prone to particular vulnerability types. We further discuss this phenomenon in ~\autoref{app:frameworks_and_vulnerability_types}.

\PP{Compare with Baselines}
Out of the 15 vulnerabilities identified by \sys, CodeQL detects 9 of them, Bandit detects 2 of them, and Vulnhuntr detects 1 of them. 
Reasons of failures mainly align with those discussed in~\autoref{ss:case-study}, such as the lack of rules in traditional static analysis tools and hallucination in LLM-based tools.

\PP{False Alarms in Detection}
\sys reports 33 findings across these projects, of which 15 we confirmed as true vulnerabilities, and 2 are less severe due to a limited attack surface, as discussed above.
All other alarms are false positives.

\sys raises more false positives than in the previous evaluations. We attribute this to these projects being more complex and generally larger than those in the previous datasets, which makes it more challenging for \sys to analyze them effectively.
These false alarms are raised align with the reasons discussed in~\autoref{ss:case-study} (i.e., hallucination and uncaught interactions). 
For example, the Flask \cc{flask.send\_from\_directory} sends a file from a specified directory but is identified by \sys as a dangerous sink that may lead to an XSS.
Another example is that the LLM incorrectly believe a username is a sensitive field that should not be directly stored in the database. It raises a false alarm for the cleartext-storage-of-sensitive-information vulnerability.

\begin{figure}[t]
    \centering
    \includegraphics[width=.45\textwidth]{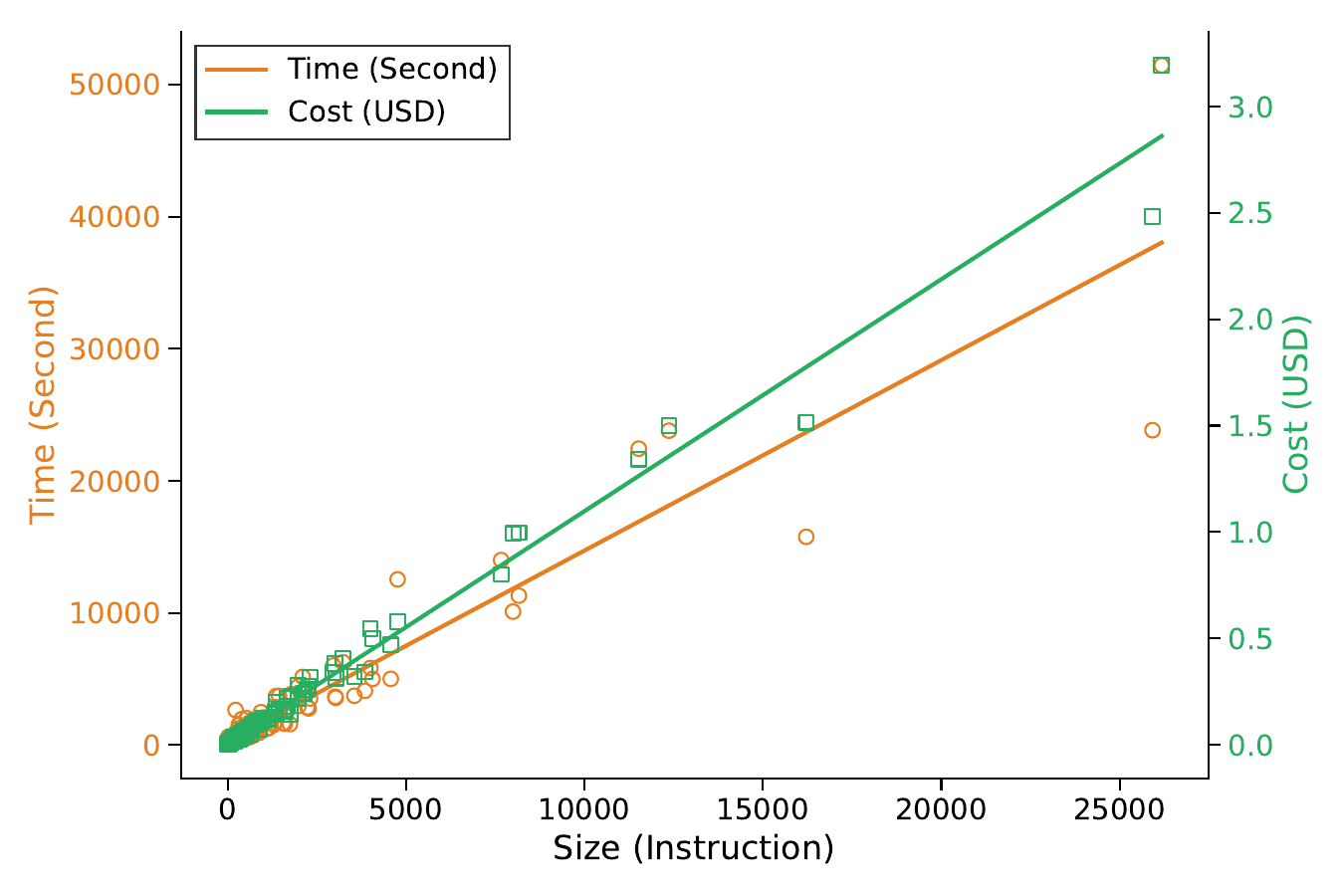}
    \caption{Time (seconds), monetary costs (USD), and the sum of LoC of analyzed functions on popular GitHub projects.}
    \label{fig:costs}
\end{figure}

\PP{Costs and Program Characteristics}
We analyze the time and monetary costs of running \sys on popular GitHub projects.
In~\autoref{tab:eval_real} we present the average time, cost, and other key metrics of running \sys on these projects, grouped by framework.
On average, a program has 15.66 endpoints and takes 3,250.75 seconds (54 minutes) and costs \$0.23 to analyze.
The preparation phase (\WC{1} in~\autoref{fig:pipeline}) takes 419.35 seconds (7 minutes) on average, accounting for 12.9\% of the total time.
The remainder is spent on LLM-assisted analysis, which makes on average 1,168.93 calls to infer \type and 523.83 calls to check security properties.

Per endpoint, analysis takes 223.34 seconds (3.7 minutes) and costs 1.64 US cents on average.
Each endpoint involves analyzing 145.90 bytecode instructions on average, requiring 81.49 LLM queries to infer \type and 36.03 LLM queries to check security properties.

To better understand the relationship between time, cost, and program characteristics, we illustrate these metrics for each project in~\autoref{fig:costs}.
In the figure, we can observe that the cost is generally proportional to the number of bytecodes analyzed, as more code requires more LLM queries, leading to higher costs.
Alternatively, while execution time generally increases with the number of bytecodes analyzed, its correlation with bytecode count is weaker than for monetary cost because program structure and complexity also significantly affect runtime.
In our experiments, the largest project analyzed has 195 endpoints and passes roughly 26k bytecode instructions to the LLM. This takes 51,454 seconds ($\approx$14 hours), and costs \$3.19.

Among frameworks, Django projects require more total analysis time but less time per endpoint than Flask and FastAPI. This likely reflects that Django apps are larger and more complex overall, whereas individual endpoints tend to be simpler.

\section{Limitations}
\label{s:limitations}

While \sys presents a novel approach to achieve \type inference and vulnerability detection, it has several limitations that need to be addressed in future work.

\PP{Statically Unknown Code} 
\sys is designed to analyze Python programs statically. However, certain dynamic features of Python may lead to uncaught interactions during the analysis. For example, if a program dynamically loads a module at runtime, \sys cannot accurately infer the types of variables associated with that module. Similarly, if the indirect call map constructed by pytype is incomplete, \sys may miss certain function calls that could affect the type inference process, and code paths involving these calls will not be analyzed.

Furthermore, \sys only analyzes code belonging to the target project and limits itself to considering only the interfaces of code in dependent libraries. However, if dependent libraries trigger certain behaviors of the target project (e.g., through class inheritance or callback functions), \sys may miss these interactions, leading to incomplete or incorrect analysis results.
The case in~\autoref{sss:missing-detections}, is an example.

\PP{Interact with Other Languages}
Many Python applications interact with other languages and runtimes (e.g., HTML templates), and vulnerabilities may arise from these cross-language interactions.
For example, when a Python function renders an HTML template that interpolates user input, XSS may arise in the generated HTML even though the Python-level value appears harmless, and vice versa. 
\sys currently offers only heuristic handling of these cross-language interactions (e.g., simple string-pattern checks and treating template render calls as opaque), which is insufficient to model language-specific escaping. A more robust solution would require cross-language analysis that accurately models the semantics and security properties of each involved language, which is an important direction for future work.

\PP{Implementation Limitations}
\sys is intended as a prototype to demonstrate the effectiveness of \type in detecting vulnerabilities. As such, it has several implementation limitations that should be addressed in future work. First, \sys currently targets Python 3.10 only: bytecode and VM behavior differ across Python releases, so additional engineering is required to support other versions reliably. Second, the implementation covers a restricted set of language constructs and libraries; advanced or highly dynamic features (e.g., metaclasses, complex decorators, and dynamic imports) may not be modeled accurately. Third, our vulnerability checks focus on a limited set of semantic rules commonly observed in web applications; broader CWE coverage and finer-grained checks are needed for wider applicability. 

\section{Related Works}
\label{s:related}

\subsection{Traditional Vulnerability Detection}
\label{ss:traditional_vuln_analysis}

Static analysis tools for vulnerability detection have long relied on manually crafted rules and pattern matching to flag risky code constructs. Targeting on this, prior works can be grouped into two complementary directions: 

\PP{Extending Search Space}
One line of work targets on expanding the expressiveness of vulnerability specifications to capture richer program semantics through previously unexploited relationships.
Examples include inter-procedural specifications that relate allocation/use/free lifecycles~\cite{zhang2025statically}, precise detection of multiple-read accesses~\cite{xu2018precise}, and leveraging class inheritance~\cite{lin2025uncovering}. 

\PP{Automatic Pattern Inference}
To reduce manual effort, a second line of work targets on deriving patterns automatically from data. Approaches include mining patches and change histories to infer repair-driven specifications~\cite{chen2025seal}, assuming normative behavior by majority voting over observed uses and flagging deviations~\cite{min2015cross}, and detecting anomalies by cross-checking functions with similar intent across a codebase~\cite{lu2019detecting}.


\subsection{Learning-Based Vulnerability Detection}
\label{ss:dl_vuln_analysis}

Code Language Models (CodeLMs) are specialized language models trained on large corpora of source code, enabling them to understand programming languages and code structures. Examples of these models include CodeBERT~\cite{feng2020codebert}, CodeLlama~\cite{roziere2023code}, and DeepSeek-Coder~\cite{deepseek_coder_2024}. However, recent works have shown that these models struggle to reliably identify and reason about security vulnerabilities in large codebases and for uncommon vulnerability types~\cite{ullah2024llms}.

Alternatively, researchers have trained models specifically for vulnerability detection. At different levels of detection granularity (e.g., repository-level~\cite{wang2024reposvul}, function-level~\cite{chakraborty2021deep,halder2025funcvul}, slice-level~\cite{cheng2021deepwukong}, and line-level~\cite{fu2022linevul}), these methods train predictive models to identify whether a code snippet is vulnerable. These approaches can autonomously detect vulnerabilities without manually defined rules, but they suffer from limited interpretability and require large labeled datasets for effective training and generalization.

\subsection{LLM-based Contextual Bug Analysis}
\label{ss:llm_contextual_analysis}

LLMs have been leveraged into the vulnerability analysis domain in two main ways:

\PP{LLMs for Direct Vulnerability Detection}
Recent works have explored using LLMs directly for vulnerability detection by leveraging their understanding of security concepts. For instance, researchers have used CWE descriptions to guide LLMs in identifying potential vulnerabilities~\cite{li2025automated} and locating vulnerable functions~\cite{wu2024effective}. Others have focused on detecting inconsistencies between code and comments to reveal potential security issues~\cite{rong2025code}. There also exists frameworks that employ multiple LLM agents working collaboratively to reduce false positives~\cite{wen2024collaboration}.

\PP{LLMs for Assisting Traditional Analysis}
LLMs have been particularly effective in identifying data flows and critical functions. Recent studies have shown considerable success in using LLMs to assist taint analysis by identifying source and sink functions~\cite{li_iris_2025,ji2025artemis} and to map handlers to call sites in embedded systems~\cite{zhao2024leveraging}.

LLMs have also been used to improve the precision of vulnerability detection. Several approaches have been proposed to reduce false positives through sophisticated analysis techniques. Some methods use LLMs to track execution constraints of variables~\cite{li2024enhancing,repoaudit2025,lekssays2025llmxcpg}, while others adopts LLMs to analyze broader code context~\cite{wen2024automatically,xia2025beyond}.

\section{Concluding Discussion}
\label{s:conclusion}

While LLMs bring powerful pattern recognition and generative capabilities, they are not a silver bullet: they can produce hallucinations and are constrained by context length. Relying solely on LLMs for program analysis can lead to unreliable results.
Conversely, static analysis offers complementary guarantees, such as soundness and provable absence of certain classes of bugs, but is constrained by a limited ability to understand program semantics at a higher level and to infer developer intent. 

Taken together, these complementary weaknesses suggest a hybrid direction. A middle ground that leverages the strengths of LLMs for reasoning and summarization, together with static analysis for analysis driven rigor, could yield more reliable and precise vulnerability detection.

As a step toward this middle ground,
we propose \type that associates both syntax and semantic types with program variables, enabling more precise type checking that can identify semantically inconsistent operations despite their syntactic correctness.
An instantiation of \type,
\sys tailored towards Python programs,
orchestrates model queries and analysis passes, employs caching and summarization to mitigate context-length limitations, and uses iterative verification loops to reduce hallucinations and improve precision. Evaluation shows that it achieves notable detection rates, demonstrating the viability of this hybrid approach.
We believe combining LLM-driven reasoning with rigorous static analysis offers a promising path toward more effective vulnerability detection in real-world software.



\bibliographystyle{plainurl}
\bibliography{p,sslab,conf}

\appendix

\section{Semantic Type System}
\label{app:type-system}

In this section, we provide a formal description of the \type system.
To start with, we define the \type of a variable or an expression $e$ as $\tau$ in the typing environment $\Gamma$, denoted as $\Gamma \vdash e : \tau$. Here, we define the following rules:

When a variable $x$ is first encountered, we initialize its \type $\tau_x$ based on its non-executable context (e.g., variable name and comments) using a function $\mathit{inf}(n_x)$, where $n_x$ is the non-executable context of $x$.

    \[
    \frac{x \notin Dom(\Gamma) \quad \tau_x = \mathit{inf}(n_x)}
        {\Gamma \vdash x : \tau_x}
    \]

After initialization, the \type of $x$ is retrieved from the typing environment $\Gamma$:
    \[
    \frac{x \in Dom(\Gamma) }
        {\Gamma \vdash x : \tau_x}
    \] 

For an operation $o$ with operands $e_1, \ldots, e_n$, if each operand $e_i$ has the \type $\tau_{e_i}$, then the operation has the \type $\tau_o$:
    \[
    \frac{\Gamma \vdash e_1 : \tau_{e_1} \quad  \ldots \quad \Gamma \vdash e_n : \tau_{e_n} \quad \tau_o = \mathit{op}(o, \tau_{e_1}, \ldots, \tau_{e_n})}
        {\Gamma \vdash o(e_1, \ldots, e_n) : \tau_o}
    \]
When analyzing an assignment statement of the form $x \leftarrow e$, where \cc{x} is a variable and \cc{e} is an expression.
We notate the variable $x$ before the assignment as $\hat{x}$.

We first infer the \type $\tau$ of the expression \cc{e} under the current typing environment $\Gamma$, i.e., $\Gamma \vdash e : \tau_e$. The assignment updates the \type of $x$ in $\Gamma$ by unifying its existing \type $\tau_x$ with the inferred \type $\tau$:
\[
\frac{\Gamma \vdash e : \tau_e \quad \Gamma(\hat{x}) = \tau_x \quad \Gamma (x) = \mathit{unify}(\tau_x,\tau_e)}
    {\Gamma \vdash x \leftarrow e}
\]

Notably, in the above rules, the function $\mathit{inf}$, $\mathit{op}$, and $\mathit{unify}$ are not formally defined. 
Indeed, these functions are implemented using LLMs in \sys, as described in~\autoref{s:design}.
Type checking is performed within the $\mathit{op}$ and $\mathit{unify}$ functions to ensure that the operations are semantically valid.

\section{Targeted CWE List}
\label{app:taret-cwes}
\begin{itemize}[noitemsep]
    \item CWE-22: Path Traversal      
    \item CWE-78: OS Command Injection    
    \item CWE-79 (CWE-80): Cross-Site Scripting (XSS)  
    \item CWE-89: SQL Injection
    \item CWE-200: Exposure of Sensitive Information to an Unauthorized Actor
    \item CWE-312: Cleartext Storage of Sensitive Information
    \item CWE-601: Open Redirect
    \item CWE-617: Reachable Assertion
    \item CWE-200: Information Exposure
    \item CWE-807: Reliance on Untrusted Inputs in a Security Decision
    \item CWE-918: Server-Side Request Forgery (SSRF)
\end{itemize}

\section{Dataset Constrution Details}
\label{app:popular-projects}

\PP{Construction of Recent CVE Dataset}
We following: 1 entry that has no CWE assigned, 4 are due to the usage of vulnerable third-party libraries, 5 entries that have the vulnerability level set to low, 5 entries that are not open-sourced in GitHub, 2 entries that focus on denial-of-service vulnerabilities, and three servers are designed for local usage only. Out of the remaining 23 entries, we take entries with their CWE number in the CWE types covered in the SVEN dataset or appears at least twice in the dataset. We eventually have 9 entries in the dataset, acrossing five CWE types, and 5 of them have been patched by the developers.

\PP{Constrution of Popular Projects Dataset} We construct a dataset of popular open-source projects that use FastAPI, Django, or Flask frameworks.

As of July 31th, 2025, we collected 37535 FastAPI projects, 30385 Django projects and 14141 Flask projects. Out of these repositories, 4734, 2440, and 2262 projects are public and has least one star. Among these projects, we select the top-100 starred projects for each framework. 

Among these projects, we manually filter out the following:
{\begin{itemize}[noitemsep,topsep=1pt]
    \item 144 projects that do not indeed use the framework in the code. This is because some projects only list the framework in the dependency file but do not actually use it in the code, or an older version of the project uses the framework but the latest version does not;
    \item 19 projects that are framework-plugins or extensions;
    \item 22 projects that are tutorials or the collection of code snippets;
    \item 9 projects that uses a framework that is built on top of the target framework (e.g., \footnote{\url{https://github.com/jlowin/fastmcp}}{\cc{fastmcp}}), and
    \item 3 projects that are duplicated repositories.
\end{itemize}

\section{Insufficient Patch}
\label{app:insufficient-patches}

In certain cases, the developer's patch may not completely fix the vulnerability, leading to repeated detections by \sys. For example, in CVE-2025-6776~\cite{CVE_2025_6776}, multiple path traversal vulnerabilities are found where the user input is used to construct file paths without proper sanitization. The developer's patch is to restrict the access to the endpoints to logged-in users. However, the underlying issue of unsanitized user input is not addressed in the patch. As a result, both CodeQL and \sys still report the vulnerabilities in both versions. 

\section{Uncaught Interactions Code Example}
\label{app:uncaught-interactions}

\hfill\begin{minipage}[c]{\linewidth}\vspace{0.5em}\begin{Verbatim}[commandchars=\\\{\},numbers=left,firstnumber=1,stepnumber=1,codes={\catcode`\$=3\catcode`\^=7\catcode`\_=8\relax}]
\PY{k}{class} \PY{n+nc}{ResendEmailSerializer}\PY{p}{(}\PY{n}{serializers}\PY{o}{.}\PY{n}{Serializer}\PY{p}{)}\PY{p}{:}
  \PY{n}{email} \PY{o}{=} \PY{n}{serializers}\PY{o}{.}\PY{n}{EmailField}\PY{p}{(}\PY{n}{label}\PY{o}{=}\PY{p}{(}\PY{l+s+s2}{\PYZdq{}}\PY{l+s+s2}{Email}\PY{l+s+s2}{\PYZdq{}}\PY{p}{)}\PY{p}{)}

  \PY{k}{def} \PY{n+nf}{validate}\PY{p}{(}\PY{n+nb+bp}{self}\PY{p}{,} \PY{n}{attrs}\PY{p}{)}\PY{p}{:}
    \PY{n}{user} \PY{o}{=} \PY{n}{get\PYZus{}object\PYZus{}or\PYZus{}404}\PY{p}{(}\PY{n}{User}\PY{p}{,} \PY{n}{email}\PY{o}{=}\PY{n}{attrs}\PY{o}{.}\PY{n}{get}\PY{p}{(}\PY{l+s+s2}{\PYZdq{}}\PY{l+s+s2}{email}\PY{l+s+s2}{\PYZdq{}}\PY{p}{)}\PY{p}{)}
    \PY{n}{attrs}\PY{p}{[}\PY{l+s+s2}{\PYZdq{}}\PY{l+s+s2}{user}\PY{l+s+s2}{\PYZdq{}}\PY{p}{]} \PY{o}{=} \PY{n}{user}
    \PY{k}{return} \PY{n}{attrs}

\PY{k}{class} \PY{n+nc}{ResendEmailView}\PY{p}{(}\PY{n}{GenericAPIView}\PY{p}{)}\PY{p}{:}
  \PY{n}{serializer\PYZus{}class} \PY{o}{=} \PY{n}{ResendEmailSerializer}

  \PY{k}{def} \PY{n+nf}{post}\PY{p}{(}\PY{n+nb+bp}{self}\PY{p}{,} \PY{n}{request}\PY{p}{,} \PY{o}{*}\PY{n}{args}\PY{p}{,} \PY{o}{*}\PY{o}{*}\PY{n}{kwargs}\PY{p}{)}\PY{p}{:}
    \PY{n}{serializer} \PY{o}{=} \PY{n+nb+bp}{self}\PY{o}{.}\PY{n}{serializer\PYZus{}class}\PY{p}{(}\PY{n}{data}\PY{o}{=}\PY{n}{request}\PY{o}{.}\PY{n}{data}\PY{p}{)}
    \PY{n}{serializer}\PY{o}{.}\PY{n}{is\PYZus{}valid}\PY{p}{(}\PY{n}{raise\PYZus{}exception}\PY{o}{=}\PY{k+kc}{True}\PY{p}{)}
    \PY{n}{user} \PY{o}{=} \PY{n}{serializer}\PY{o}{.}\PY{n}{validated\PYZus{}data}\PY{p}{[}\PY{l+s+s2}{\PYZdq{}}\PY{l+s+s2}{user}\PY{l+s+s2}{\PYZdq{}}\PY{p}{]}
    \PY{k}{if} \PY{n}{user}\PY{o}{.}\PY{n}{is\PYZus{}verified}\PY{p}{:}
      \PY{k}{return} \PY{n}{Response}\PY{p}{(}\PY{p}{\PYZob{}}\PY{l+s+s2}{\PYZdq{}}\PY{l+s+s2}{detail}\PY{l+s+s2}{\PYZdq{}}\PY{p}{:} \PY{l+s+s2}{\PYZdq{}}\PY{l+s+s2}{...}\PY{l+s+s2}{\PYZdq{}}\PY{p}{\PYZcb{}}\PY{p}{)}
\end{Verbatim}\end{minipage}\vspace{0.5em}

Here, we present a code example that illustrates the uncaught interactions during \sys's analysis. The example is taken from the \footnote{\url{https://github.com/MaryamKalantarii/MiniKala}}{MiniKala} project in the explored popular projects dataset. In this example, the \cc{ResendEmailView} class uses a serializer \cc{ResendEmailSerializer} for validating the email address provided by the user. 
However, in the \cc{post} method of the \cc{ResendEmailView} class (line 13), the called \cc{is\_valid} (line 15) and \cc{validated\_data} (line 16) are not defined in the \cc{ResendEmailSerializer} class. Instead, these methods are inherited from the parent class \cc{django.serializers.Serializer}. 
Similarly, the \cc{validate} method (line 4) in the \cc{ResendEmailSerializer} class is also not called.
However, this function will be invoked during the execution of the \cc{is\_valid} method in the parent class to validate the email field.

During \sys's analysis, it does not explore this inheritance relationship and thus misses this interaction. As a result, \sys cannot infer the types of the variables \cc{email} and \cc{validated\_data['email']} accurately, leading to incomplete analysis results.

\section{Frameworks and Vulnerability Types}
\label{app:frameworks_and_vulnerability_types}

While our work focuses on detection of vulnerabilities in the three frameworks, we observe that certain types of vulnerabilities are more prevalent in one framework compared to others. 

\PP{Open Redirect in Django} In Django, open redirect vulnerabilities are particularly common. This is primarily due to the framework's use of the ``next'' parameter in its authentication views, which allows redirection to a specified URL after login. One potential reason for the prevalence of open redirect vulnerabilities in Django applications is the common mentioning of using the ``next'' parameter in various tutorials and documentation~\cite{geeks}. This widespread practice may lead developers to inadvertently introduce open redirect vulnerabilities if they do not properly validate the ``next'' parameter before using it for redirection. Alternatively, a similar design pattern is not commonly used in Flask or FastAPI, which may explain the lower incidence of open redirect vulnerabilities in applications built with these frameworks.

\PP{File and Data Management} In Flask and FastAPI, we observe a higher occurrence of file and database management vulnerabilities, such as arbitrary file read and SQL injection. This could be attributed to the fact that both frameworks provide more flexibility and commonly allows the user to choose their own libraries for handling file uploads and database interactions. This flexibility, while beneficial for customization, may lead to inconsistent security practices among developers, increasing the likelihood of vulnerabilities in these areas. Alternatively, Django's built-in Object-Relational Mapper (ORM) avoids the need for raw SQL queries or filesystem interactions, which may contribute to a lower incidence of such vulnerabilities in Django applications.

\end{document}